\documentclass{article}
\newcommand{\beq}{\begin{equation}}
\newcommand{\eeq}{\end{equation}}
\newcommand{\beqa}{\begin{eqnarray}}
\newcommand{\eeqa}{\end{eqnarray}}

\newcommand{\diag}{{\mathrm{diag}}}
\newcommand{\mpl}{M_{Pl} }
\newcommand{\wtree}{W_{\mathrm{tree}}}
\newcommand{\luv}{\Lambda_{\mathrm{UV}} }
\newcommand{\guv}{g_{\mathrm{UV}} }
\begin{document}

\begin{titlepage}
\begin{flushright}
\end{flushright}

\vspace{15pt}

\begin{center}

{\Large{\bf  Supersymmetry Breaking}\footnote{Lectures given at the 
Les Houches Summer School (Session LXXXIV) on ``Particle Physics Beyond the
Standard Model'', Les Houches, France, August 1-26, 2005.}
}

\vspace{15pt}

{
{Yael Shadmi}
}

\vspace{7pt}
{\small
{\it{Physics Department, Technion---Israel Institute of Technology,
Haifa 32000, Israel \\} }
\vspace{4pt}
yshadmi@physics.technion.ac.il
}
\end{center}

\begin{abstract}
\noindent
These lectures provide a simple introduction to supersymmetry breaking.
After presenting the basics of the subject and illustrating them 
in tree-level examples, we discuss dynamical supersymmetry breaking,
emphasizing the role of holomorphy and symmetries in restricting 
dynamically-generated superpotentials.
We then turn to mechanisms for generating the MSSM supersymmetry-breaking 
terms, including ``gravity mediation'', gauge mediation, and 
anomaly mediation. 
We clarify some confusions regarding the decoupling of heavy fields 
in general and D-terms in particular in models of anomaly-mediation. 
\end{abstract}

\end{titlepage}

\section{Introduction}
Need we motivate lectures on supersymmetry breaking?
Not really. If there is supersymmetry in Nature,
it must be broken. But it's worth emphasizing
that the breaking of supersymmetry, namely, the masses
of superpartners, determines the way supersymmetry would
manifest itself in experiment.

From a purely theoretical point of view, supersymmetry
breaking is a very beautiful subject, and I hope these
lectures will convey some of this beauty.

It is very hard to cover supersymmetry-breaking
in three lectures. 
In the first lecture, section~\ref{basic}, we will describe the essentials
of supersymmetry breaking.
In the second lecture, section~\ref{beyond}, we will study dynamical
supersymmetry breaking.
In the last lecture, section~4, we will describe several mechanisms
for generating supersymmetry-breaking terms for the standard-model
superpartners.
This section can be read independently of section~\ref{beyond}.

For lack of time, we will not cover supersymmetry-breaking mechanisms,
or mechanisms for mediating the breaking, that rely on
extra dimensions (we will discuss anomaly-mediation, 
because it is always present in four dimensions).

These lectures assume basic knowledge of supersymmetry
(essentially the first seven chapters of 
Wess and Bagger\cite{Wess:1992cp},
whose notations we will use).
I tried to make section~\ref{beyond} self-contained,
but a serious treatment of non-perturbative effects
in supersymmetric gauge theories is beyond the scope
of these lectures. 
For excellent reviews of the subject see, e.g., 
\cite{Intriligator:1995au, Shifman:1999mv, terningtasi}.
For more details and examples of dynamical supersymmetry breaking, 
see~\cite{Shadmi:1999jy,Poppitz:1998vd}.
Finally, ref.~\cite{Giudice:1998bp} is a comprehensive review
of gauge-mediation models.
\section{Basic features of supersymmetry breaking}
\label{basic}
In this section, we will discuss the fundamentals of supersymmetry
breaking: the order parameters for the breaking, the Goldstone
fermion, $F$-type and $D$-type tree-level breaking, and some general
criteria for determining when supersymmetry is broken.
The discussion will mostly be in the framework of  ${\cal N}=1$ 
global supersymmetry,
but we will end this section by commenting on how things are modified
for  local supersymmetry.

\subsection{Order parameters for supersymmetry breaking}
\label{order}
When looking for spontaneous supersymmetry breaking,
we are asking whether the variation of some field
under the supersymmetry transformations is non-zero
in the ground state,
\beq
\langle 0\vert\delta(\mathrm{field})\vert 0\rangle \neq 0\ .
\eeq
For a chiral superfield $\phi$, with scalar component $\tilde\phi$,
fermion component $\psi$, and auxiliary component $F$,
the supersymmetry variation are roughly (omitting numerical coefficients),
\beqa\label{var}
\delta_\xi\tilde\phi(x)&\sim& \xi \psi(x) \nonumber\\
\delta_\xi\psi(x) &\sim& i\sigma^\mu\bar\xi \, \partial_\mu\tilde\phi(x)
+\xi F(x)\\
\delta_\xi F(x)&\sim& i \bar\xi \bar\sigma^\mu 
\partial_\mu \psi(x) \ ,\nonumber
\eeqa
where $\xi$ parameterizes the supersymmetry variation.
Clearly, the only Lorentz invariant on the RHS of eqn.~(\ref{var})
is $F$,
so supersymmetry is broken if 
\beq\label{fterm}
<F>\neq 0\ ,
\eeq
and the field whose variation is non-zero in this case is
the fermion, 
$\langle 0\vert\delta_\xi\psi(x)\vert 0\rangle \neq 0$.

Similarly, for the vector superfield, 
only the gaugino variation can be non-zero 
\beq
\langle 0\vert\delta_\xi\lambda(x)\vert 0\rangle \propto
\langle 0\vert D\vert0\rangle \neq 0\ ,
\eeq
so a non-zero $\langle D\rangle$ signals supersymmetry breaking.

A much more physical order parameter for global supersymmetry breaking
is the vacuum energy.
The supersymmetry algebra contains the translation operator $P_\mu$
\beq
\{Q_\alpha,\bar Q_{\dot\alpha}\} = 2\sigma^\mu_{\alpha\dot{\alpha}}\, P_\mu\ ,
\eeq
where $Q$ is the supersymmetry generator.
Therefore the Hamiltonian $H$ can be written as
\beq
H ={1\over4}\, (\bar{Q}_1 Q_1 + \bar{Q}_2 Q_2 +\mathrm{h.c.})\ .
\eeq
Since this is a positive operator, the energy of a supersymmetric
system is either positive or zero.
Furthermore, if supersymmetry is unbroken, the vacuum is annihilated by the
supersymmetry generators, and
\beq
E_{\mathrm{vacuum}}= \langle 0\vert H\vert 0\rangle =0\ .
\eeq
Thus, a non-zero vacuum energy signals spontaneous supersymmetry breaking.

In order to know whether global supersymmetry is spontaneously
broken, we therefore need to study the minima of the scalar potential, and
see whether there is a minimum with zero energy.

\subsection{The scalar potential and flat directions}
\label{sscalar}
In a theory with chiral superfields $\phi_i$, superpotential
$W(\phi_i)$ and K\"ahler potential $K(\phi_i, \phi_i^\dagger)$,
the scalar potential is given by
\beq\label{vf}
V_F= K^{-1}_{i* j} {\partial W^*\over \partial\phi_i^*} 
{\partial W\over\partial \phi_j} = K^{-1}_{i j} F_i^* F_j
\ ,
\eeq
where 
\beq
K_{ij*}= {\partial^2 K\over \partial\phi_i \partial\phi_j*}\ .
\eeq
In eqn.~(\ref{vf}) we used the fact that, on-shell, the auxiliary
fields are given by
\beq
F_i={\partial W\over\partial\phi_i}\ .
\eeq 

If there are gauge interactions in the theory the scalar potential
has additional contributions and is given by
\beq
V= V_F + V_D = V_F + {1\over 2} g^2\sum_a \left(D^a\right)^2\ ,
\eeq
where $D^a=\sum_i\phi_i^\dagger T^a \phi_i$.
As expected, the scalar potential is non-negative,
and again we see that supersymmetry is broken
by a non-zero $F$ and /or $D$ vacuum expectation value (VEV).
Only then is the ground state energy non-zero.

To look for the zeros of the scalar potential (in field space)
in a theory with gauge interactions, we need to do the following:
\begin{enumerate}
\item Find the sub-(field)space for which $D^a=0$.
This is often called the space of ``D-flat directions''.
Note that along these directions, the potential is not merely
flat, but rather zero\footnote{The reason 
why these directions are called ``flat''
will become clear once we discuss radiative corrections.
Typically, in non-supersymmetric theories, if we have a flat
potential at tree level, the degeneracy is lifted
by radiative corrections. As we will see, in supersymmetric
theories, if the ground state energy is zero at tree-level,
it remains zero to all orders in perturbation theory.
Therefore, the directions in field space for which $V=0$
are the only ones that are truly flat---they remain zero 
to all orders in perturbation theory.
}.
The space of $D$-flat directions can be parametrized
by the VEVs of the chiral gauge invariants that one can
construct from the fundamental chiral fields of the theory.
This is an extremely useful result and we will often use
it in the following.
\item If for a subspace of the $D$-flat directions we also have
$F_i=0$ (for all $F_i$'s), then the potential is zero.
The sub-(field) space for which this happens is often
called the ``moduli space''.
\end{enumerate}
To look for supersymmetry breaking, we will be interested then in the
moduli-space of the theory. 
If there is no moduli space, supersymmetry is broken.

\vspace*{0.5cm}
\noindent
{\bf Exercise: D-flat directions}: Consider an $SU(N)$ gauge theory
with chiral fields $Q_i\sim N$, $\bar{Q}^A\sim \bar{N}$, with 
$i, A=1,\ldots,F$. (This theory is usually called $SU(N)$ with $F$ flavors.)
Assume $F<N$. Denote the $SU(N)$ gauge index by $\alpha$.
Show that 
\beq\label{dflat}
Q_{i \alpha}= \bar{Q}_{i \alpha} = v_i \delta_{i \alpha} \ ,
\eeq 
are $D$-flat.
The $D$-flat directions of the theory are then given by~(\ref{dflat})
up to global $SU(F)_L\times SU(F)_R$ and gauge rotations.

As mentioned above, an alternative parameterization of the $D$-flat
directions is in terms of the VEVs of the gauge invariants
of the theory.
In this case, the only chiral gauge invariants are the ``mesons''
$M_i^A= Q_i\cdot Q^A$. 
Indeed, using the global symmetry we can always write
the meson VEVs as 
\beq
M_i^A= \diag(V_1,V_2,\ldots, V_N)\ .
\eeq
and the two parameterizations are clearly equivalent
$V_i \leftrightarrow v_i^2$.

\subsection{The Goldstino}
\label{goldstino}
With broken supersymmetry $Q_\alpha \vert 0\rangle$ is non-zero.
What is it then? The generator of a broken bosonic global
symmetry gives the Goldstone boson. 
Likewise, $Q_\alpha \vert 0\rangle$ gives the Goldstone fermion of
supersymmetry breaking, or ``Goldstino'', which we denote by 
$\psi^G_\alpha(x)$.

To see the Goldstino concretely, we should examine the supersymmetry
current, and look for a piece that is linear in the fields.
The supersymmetry current is of the form
\beq
J^\mu_\alpha \sim \sum_\phi \, 
{\delta{\cal L}\over \delta(\partial_\mu\phi)}\, 
(\delta\phi)_\alpha\ ,
\eeq
where $\delta\phi$ is the supersymmetry variation of the 
the field $\phi$.
Since
${\delta{\cal L}\over \delta(\partial_\mu\phi)}$ cannot get a VEV,
a term that is linear in the fields can only occur when 
$\delta\phi$ gets a non-zero VEV.
As we saw before, the only fields whose supersymmetry variations
can have non-zero VEVS are the fermion of the chiral superfield,
$\psi$ (the VEV of whose variation is $F$), and the fermion
of the vector superfield, $\lambda$ (the VEV of whose variation is $D$).
Thus,
\beq
J_\mu^\alpha\sim 
\sum_i {\delta{\cal L}\over 
\delta(\partial_\mu\psi_{i\alpha})}\,\langle F_i\rangle +
{1\over\sqrt2} \sum_a {\delta{\cal L}\over 
\delta(\partial_\mu\lambda_\alpha^a)}\,\langle D^a\rangle\ ,
\eeq
so that
\beq
\psi_\mu^G\sim \sum_i \langle F_i\rangle \psi_i 
+ \sum_a \langle D^a\rangle \lambda^a\ .
\eeq
We see that the Goldstino is a combination of the fermions
that correspond to non-zero auxiliary field VEVs.

To demonstrate the basics we have seen so far, 
let us now turn to two examples of supersymmetry breaking.
These examples will also illustrate some other
general features of supersymmetry breaking.

\subsection{Tree-level breaking: $F$-type}
\label{ftype}
In this section we will study a variation
of the O'Raifeartaigh model~\cite{O'Raifeartaigh:1975pr}, 
with chiral fields $Y_i$, $Z_i$,
and $X$ with $i=1,2$,
with the superpotential 
\beq\label{orafw}
W = X(Y_1 Y_2 -M^2) + m_1 Z_1 Y_1 + m_2 Z_2 Y_2\ ,
\eeq
where $M$ and $m_i$ are parameters with the dimension of mass.
Note that the superpotential has a term that is linear in one of the
fields ($X$). This is crucial for breaking supersymmetry at tree-level.

The original O'Raifeartaigh model is obtained by identifying
$Y_1=Y_2=Y$, and $X_1=X_2=X$. 
We are complicating the model in order to illustrate
the interplay between broken global symmetries
and supersymmetry breaking, which we will get to
later.
But let's postpone that, and see whether the model breaks supersymmetry.

Since there are no gauge interactions in the model, we don't
have to worry about $D$-terms, and we can turn directly
to finding whether there are $F$-flat directions for which
the potential vanishes.
Equating all the $F$-terms to zero we have the following equations:
\beqa
&1&\ \ \ Y_1 Y_2 = M^2\ \ \ \ \  (F_X)\nonumber \\
&2& \ \ \ X Y_2 + m_1 Z_1 =0\ \ \ \ \ \ (F_{Y_1})\nonumber \\
&3& \ \ \ X Y_1 + m_2 Z_2 =0\ \ \ \ \ \ (F_{Y_2}) \nonumber \\
&4& \ \ \ m_1 Y_1 = 0 \ \ \ \ \ \  (F_{Z_1})  \nonumber\\
&5& \ \ \ m_2 Y_2 = 0 \ \ \ \ \ \ (F_{Z_2}) \nonumber
\eeqa
Clearly, equations~4 and~5 clash with equation~1.
There is no point for which the potential vanishes,
and supersymmetry is broken.
Note that it is crucial that $M$, $m_1$ and $m_2$ are
all non-zero.
If $M=0$, there is no linear term in the superpotential,
and the origin of field space is always a supersymmetric point\footnote{This
will no longer hold when we discuss non-perturbative effects,
which can give superpotential terms with negative powers of the fields.}.
If for example, $m_2=0$, we can have a solution with $Y_1\rightarrow 0$
and $Y_2\rightarrow\infty$, such that their product is $M^2$.

You may be gasping with disbelief at how simple
supersymmetry breaking is. 
And it's true: given a superpotential, finding out
whether supersymmetry is broken simply amounts to
solving a system of equations.
The tricky part, as we will see, is to derive the superpotential,
which usually involves understanding the dynamics of the theory.

As we saw above, supersymmetry is broken in this model.
Very often, this is all one can say about a model.
There are many other questions one can ask, such as: 
Where is the minimum of the potential?
Which global symmetries are preserved in this minimum?
What is the ground state energy? What is the light spectrum?
To answer these questions, we need to know the K\"ahler potential
of the theory.

In fact, we have already made an implicit assumption about the
K\"ahler potential when we determined that supersymmetry is
broken. 
We found that some $F$ terms are non-zero in the model,
but inspecting~(\ref{vf}), we see that the potential can still
vanish if $K_{ij}$ blows up. 
So we are assuming that the K\"ahler potential is well behaved.
For the simple chiral model we wrote above, this is a completely
innocent assumption. 
But in general, when we study gauge theories with complicated
dynamics, this is an important caveat to keep in mind.

But let's take the tree-level K\"ahler potential of our
toy model to be canonical.
The potential is then
\beq
V= \left\vert Y_1 Y_2 - M^2\right\vert^2 + 
\left[ \left\vert X Y_2+ m_1 Z_1 \right\vert^2 + 
m_1^2 \left\vert Y_1^2\right\vert^2
+ 1\leftrightarrow2 \right]\ .
\eeq

\noindent
{\bf Exercise:} Show that for $m_, m_2\ll M$, the potential is
minimized along
\beqa
\langle Y_1 \rangle &=& v_1 \equiv {\sqrt {m_2\over m_1}}\, 
{\sqrt{M^2-m_1 m_2}} \nonumber\\
 Z_1&=&-{1\over m_1}\, X Y_2\ ,
\eeqa
and similarly for $1\leftrightarrow2$.

Instead of an isolated minimum,
there is a direction in field space for which
$V$ is constant and non-zero.
This is typical of O'Raifeartaigh like models.
The degeneracy is removed at the loop level.
For example, in our toy model the true minimum
will occur at $X=Z_i=0$.

We now turn to a useful criterion for supersymmetry 
breaking~\cite{Affleck:1983vc, Affleck:1984xz}.
Suppose a theory has
\begin{enumerate}
\item A spontaneously broken global symmetry
\item No classical flat directions 
\end{enumerate}
then supersymmetry is broken.

Let us illustrate this in our toy model.
As we saw above, the model has no flat directions.
Furthermore, there is a $U(1)$ global symmetry, under which
we can choose the charges to be 
\beq
X(0)\ \ Y_1(1)\ \ Y_2(-1)\ \ Z_1(-1)\ \ Z_2(1)\ .
\eeq
Take for simplicity $m_1=m_2=m <<M$. 
The ground state is at  
\beq
\langle Y_1\rangle = \langle Y_2\rangle =v = \sqrt{M^2-m^2}\ .
\eeq
So the $U(1)$ is broken and there is a massless Goldstone
boson, which we can parameterize as $\phi_R$ with
\beqa
Y_1 &=& v e^{i(\phi_R+i\phi_I)}\nonumber\\
Y_2 &=& v e^{-i(\phi_R+i\phi_I)}\ .
\eeqa  
Consider the potential at $X=Z_i=0$,
\beq
V= \left\vert Y_1 Y_2 -M^2\right\vert^2 
+ m^2 \, \left(\left\vert Y_1\right\vert^2 
+ \left\vert Y_2\right\vert^2\right)\ .
\eeq
As expected, $\phi_R$ drops out, but $\phi_I$ doesn't.
However, for the supersymmetric theory with $m=0$, $\phi_I$ drops
out too.
What we are seeing of course is that the supersymmetric theory
is invariant under the ``complexified'' $U(1)$.
With unbroken supersymmetry, the massless Goldstone boson is
part of a massless chiral superfield, so there must be an
{\it additional} massless real scalar, and together they form
a complex scalar.
In our example, the Goldstone is $\phi_R$, and it corresponds
to a compact flat direction.
In the supersymmetric theory ($m=0$), the Goldstone is accompanied
by another massless  scalar, $\phi_I$, which corresponds to
a non-compact flat direction. 
When $m\neq0$, there is no non-compact flat direction and therefore
no other massless scalar.
Thus, the Goldstone cannot be part of a supersymmetric multiplet
and supersymmetry must be broken.

In our toy example, it was easy to verify directly that supersymmetry
is broken.
But in some examples, where a direct analysis is impossible,
it is still possible to show that there are no classical flat
directions, and that the global symmetry of the model
is broken, and thus to conclude that supersymmetry is broken.
We will see such an example in the next lecture.

\subsection{Tree-level breaking: $D$-type}
\label{dtype}
In this section we will study the Fayet-Iliopulos model~\cite{Fayet:1974jb},
in which supersymmetry is broken by a non-zero $D$ term (and/or $F$ term).
The model has a $U(1)$ gauge symmetry.
The important observation is that the auxiliary field
of the $U(1)$ vector field is gauge invariant,
and therefore can appear in the Lagrangian.
(From the point of view of supersymmetry,
we can always add an auxiliary field to the Lagrangian
because its supersymmetry variation is a total derivative.)
Consider then a model with chiral superfields $Q$ and $\bar{Q}$,
whose $U(1)$ charges are $1$ and $-1$ respectively, and with
the K\"ahler potential
\beq
K= Q^\dagger e^V Q + \bar{Q}^\dagger e^{-V} \bar{Q} + \xi_{\mathrm{FI}}\, V\ ,
\eeq
and superpotential
\beq
W= m Q \bar{Q}\ ,
\eeq
where $V$ is the vector superfield.

The potential is
\beq
V= {1\over2} g^2 \,
\left[ \vert{Q}\vert^2 -\vert{\bar{Q}}\vert^2 + \xi_{\mathrm{FI}}\right]^2
+ m^2\, \left[ \vert{Q}\vert^2 +\vert{\bar{Q}}\vert^2\right] \ .
\eeq
We see that the potential is never zero.
For the $D$-part to vanish we need $\langle{\bar{Q}}\rangle\neq0$, but
then the $F_Q$ term 
\beq
{\partial W\over \partial Q}= m\bar{Q} \neq 0\ .
\eeq

\noindent
{\bf Exercise:} Show that the minimum is at
\begin{enumerate}
\item $\langle{Q}\rangle=\langle{\bar{Q}}\rangle=0$  
for $g^2 \xi_{\mathrm{FI}} <m^2$.
In this case the $U(1)$ is unbroken, the $D$-term is non-zero,
but all $F$-terms vanish.
\item $\langle{Q}\rangle=0$, 
$\langle{\bar{Q}}\rangle=v = \sqrt{2} \sqrt{\xi_{\mathrm{FI}}-m^2/g^2}$
  for $g^2 \xi_{\mathrm{FI}}^2 > m^2$.
In this case the $U(1)$ is broken, the $D$-term is non-zero,
and one $F$-term is non-zero.
\end{enumerate}

\noindent
{\bf Exercise:} Show that the Goldstino is 
\beq
\psi_G \sim m \lambda +{i\over2} g v \psi_Q\ ,
\eeq
where $\lambda$ is the gaugino and $\psi_Q$ is the $Q$-fermion.
We explicitly see that the Goldstino is a combination
of fermion fields whose $F$- or $D$-terms are non-zero.

\subsection{Going local}
\label{local}
So far we only discussed global supersymmetry,
so let us briefly mention which parts of our discussion
above are modified when we promote supersymmetry
to a local symmetry. 
For lack of time and space, we just present here the results.
Although we can't see the origin of these results,
they are still useful in order to understand, at least
qualitatively, what Nature looks like if it has spontaneously 
broken supersymmetry.

$\bullet$ The order parameter for $F$-type breaking now becomes
\beq
D_\phi W = {\partial W\over \partial\phi}
+ {1\over \mpl^2}\, {\partial K\over \partial\phi}\, W
\ .
\eeq
If we decouple gravity by taking the Planck scale $\mpl$
to infinity, this reduces to~(\ref{fterm}).

$\bullet$
The vacuum energy is no-longer an order parameter for supersymmetry breaking.
This is very fortunate, because we certainly don't want the 
cosmological constant
to be of the order of the supersymmetry breaking scale.
The scalar potential is now (omitting $D$-terms)
\beq
V = e^{K/\mpl^2}\, \left[ (D_i W)^* K_{i j}^{-1} (D_j W) -
{3\over \mpl^2} \vert W\vert^2\right] \ .
\eeq
We can always shift the superpotential by a constant,
$W(\phi)\rightarrow W(\phi) + W_0$
so that $V=0$ even when $D_i W=0$.

$\bullet$
When supersymmetry is broken, the gravitino gets a mass.
The Goldstino is eaten by the gravitino, and supplies
the extra two degrees of freedom required for a massive
gravitino.

$\bullet$ The supergravity multiplet contains the graviton, gravitino,
and auxiliary fields.
When supersymmetry is broken, the scalar auxiliary field
of the supergravity multiplet acquires a VEV.

In the last lecture, when we discuss how supersymmetry breaking terms
are generated for the minimal supersymmetric standard model (MSSM),
we will need to know how a non-zero VEV of the supergravity scalar
auxiliary field affects the MSSM fields.
So we need to know how this auxiliary field couples to chiral and vector
fields.
It is convenient to parameterize this auxiliary field as the $F$-component
of a {\it non-dynamical} chiral superfield\footnote{This field is called 
the chiral compensator, because
it is often introduced in order to write down a superspace
Lagrangian for supergravity that is manifestly invariant under
Weyl-rescaling. Note that the lowest component of $\Phi$ breaks the
``fake'' Weyl invariance. Non-dynamical 
fields of this type, which are introduced in order to make
the Lagrangian look invariant under some fake symmetry, are called spurions.}
\beq
\Phi= 1+ F_\Phi \theta^2\ .
\eeq
The supergravity auxiliary field then couples to chiral and vector superfields
through the following rescaling of the usual Lagrangian.
\beq\label{phi}
{\cal L} = \int d^\theta \Phi^3 W(Q) 
+ \int d^4\theta \Phi^\dagger \Phi K(Q^\dagger, e^V Q)
+ \int d^2\theta \tau W^\alpha W_\alpha\ .
\eeq
It is easy to see from this that $\Phi$ is related to scale transformations.
We can also see from the Lagrangian~(\ref{phi}) that 
when supersymmetry is broken, $F_\Phi$ becomes
non-zero.
Equation~(\ref{phi}) will be our starting point when we discuss
anomaly mediated supersymmetry breaking in section~\ref{amsb}.

\section{Beyond tree level: dynamical supersymmetry breaking}
\label{beyond}
Consider a supersymmetric gauge theory with some tree-level superpotential
$W_{\mathrm{tree}}$, and with a minimum at zero energy,
$V_{\mathrm{tree}}=0$. 
Then the ground state energy remains zero to all orders
in perturbation theory~\cite{Ferrara:1974fv,Wess:1973kz,Grisaru:1979wc}.
This follows from the ``non-renormalization'' of the superpotential---
the tree-level superpotential is not corrected in perturbation theory,
which in turn, follows from the fact that the superpotential is
a holomorphic function of the fields~\cite{Seiberg:1993vc}.
We will not prove here this non-renormalization theorem, 
but we will see in detail two examples of how holomorphy
and global symmetries dictate the form of the superpotential
in section~\ref{32}.
It will be clear in these examples that the tree-level superpotential
is not corrected radiatively.

This leads to one of the most important results about supersymmetry
breaking: If supersymmetry is unbroken at tree-level, it can 
only be broken by non-perturbative effects.
Only the dynamics of the theory can generate a non-zero potential.
This makes the study of supersymmetry breaking hard
(and interesting!).
However, holomorphy, which forces us to consider non-perturbative 
phenomena when studying supersymmetry breaking,  
also comes to our aid. As we will see, we can say a lot
about the dynamics of supersymmetric theories based on holomorphy.

Before going on, let us pause to say a word about one
kind of non-perturbative phenomenon---instantons\footnote{This is 
intended for students who have never heard about instantons,
and would still like to follow these lectures.
It is by no means a serious introduction to instantons,
and I refer you to~\cite{terningtasi} for an 
introduction to instantons in supersymmetric gauge theories.},
which we will encounter in the following.
Instantons are classical solutions of the Euclidean Yang-Mills
action that approach pure gauge for $\vert x\vert \rightarrow\infty$.
Therefore, the field strength for these solutions goes to
zero at infinity, and
the instanton action is finite.
The one-instanton action is
\beq
S_{\mathrm{inst}}= {1\over2 g^2}\,
\int d^4x F_{\mu\nu}^2 \sim {8\pi^2\over g^2}\ ,
\eeq
where $g$ is the gauge coupling.
If there are fermions charged under the gauge group, instantons
can generate a fermion interaction with strength proportional
to the instanton action, $\exp(-8\pi^2/g^2)$.
The gauge coupling is of course scale-dependent,
and obeys at one-loop
\beq
\mu {dg\over d\mu} =-{b\over16\pi^2}\, g^3
\eeq
(In our conventions, ${\cal N}=1$ $SU(N)$ with $F$ flavors has
$b=3N-F$.)
So the instanton-generated interactions involve
\beq
\exp\left(-{8\pi^2\over g^2(\mu)}\right)= {\Lambda^b\over \mu^b}\ ,
\eeq
where $\Lambda$ is the strong coupling scale of the theory.

In an $SU(N)$ theory with $F=N-1$ flavors, instantons generate
fermion-scalar interactions that can be encoded by the 
superpotential~\cite{Affleck:1983mk}
\beq
W_{np}= \left(  {\Lambda^{3N-F}\over \det(Q\cdot\bar{Q}) }
\right)^{{1\over N-F}}\ .
\eeq
We will study this example in detail below.

Going back to supersymmetry breaking, we see that if the
ground state energy is zero at tree-level (unbroken supersymmetry),
only dynamical effects can alter that, and therefore
the full ground state energy, or supersymmetry-breaking scale, 
is proportional to some strong coupling scale $\Lambda$.
This has a profound implication: If a theory breaks supersymmetry
spontaneously, with supersymmetry unbroken at tree-level,
then the supersymmetry breaking scale, or the ground state energy,
is proportional to some strong coupling scale $\Lambda$,
\beq
E_{\mathrm{vac}}\sim \Lambda \sim M_{UV}\, e^{-{8\pi^2\over g^2(M_{UV})}}\ ,
\eeq
where $M_{UV}$ is the cutoff scale of the theory, say, $\mpl$.
Thus, supersymmetry can do much more than {\it stabilize}
the Planck-electroweak scale hierarchy.
It can actually {\it generate} this hierarchy
if it's broken dynamically~\cite{Witten:1981nf},
because the factor $e^{-{8\pi^2\over g^2(M_{UV})}}$ can easily be
$10^{-17}$.

In general, there are three types of (dynamical) supersymmetry-breaking
models.
\begin{enumerate}
\item In some models we can only tell that supersymmetry is broken
based on indirect arguments.
In particular, we have no information about the potential
of the theory, and all we know is that the supersymmetry-breaking scale
is of the order of the relevant strong-coupling scale.
\item In some models, we can derive the superpotential at low-energies
(in variables such that the K\"ahler potential is non-singular),
and conclude that some $F$-terms are non-zero.
Such models are often called ``non-calculable'', because apart
from determining that supersymmetry is broken, we cannot
calculate any of the properties of the ground state (including
the supersymmetry breaking scale).

How do we determine the superpotential in these models?
There are many methods, some of which we will see today.
These typically involve holomorphy, global symmetries,
known exact results and even Seiberg duality.
\item In some models, we can calculate the superpotential
as above, but for certain ranges of parameters, the theory
is weakly coupled and we can also calculate the K\"ahler potential.
Then we can compute the supersymmetry breaking scale,
the light spectrum, and other properties of the ground state.

Roughly, these models have the following behavior.
There is a tree-level superpotential $W_{tree}$, with some
couplings $\lambda$, that lifts all flat directions (classically).
Because $W_{tree}$ is a polynomial in the fields,
it vanishes at the origin, and grows for large field VEVs.
On the other hand, non-perturbative effects generate
a potential that is strong in the origin of field space,
but decreases for large field VEVs (because the gauge symmetry
is Higgsed with large scalar VEVs, so the low energy is weakly coupled).
The interplay between the tree-level potential and the
non-perturbatively generated potential may give a 
supersymmetry breaking ground state.
Clearly, if we decrease the tree-level coupling $\lambda$,
$V_{tree}$ becomes smaller, so that the ground state is
obtained at larger values of the field VEVs, where the theory
is weakly coupled.
\end{enumerate}

In the remainder of this section, we will demonstrate this through
two examples out of the many known supersymmetry breaking models.  
We will spend most of our time studying the $3-2$ model.
This example will illustrate how holomorphy, symmetries
and known results about the superpotentials of various
theories, completely determine the superpotential of the model.
\subsection{Indirect analysis---$SU(5)$ with single antisymmetric}
\label{indirect}
We will now see an example of the first type of models
discussed above, where there is only indirect evidence for 
supersymmetry breaking. 
We will apply here the criterion explained in section~\ref{ftype}:
If a theory has broken global symmetries and no flat directions,
supersymmetry is broken.
Our example is an $SU(5)$ gauge theory with fields $T\sim10$,
$\bar{F}\sim\bar5$~\cite{Affleck:1983vc,Meurice:1984ai}.
As explained above, the $D$-flat directions of a gauge theory can be 
parametrized by the chiral gauge invariants.
Since we cannot form any gauge invariants out of $T$ and $\bar{F}$,
there are no flat directions.

The global anomaly-free symmetry of the model is $G=U(1)\times U(1)_R$,
with charges $T(1,1)$ and $\bar{F}(-3,-9)$.
We can now argue, based on `t Hooft anomaly matching,
that $G$ is spontaneously broken.

So let's show that $G$ is (most likely) broken.
First, the $SU(5)$ theory probably confines.
(We stress that we cannot prove this, but since
this $SU(5)$ is asymptotically free, with few matter fields,
this is a very likely possibility.)
Suppose then that the global symmetry is unbroken.
Then the $SU(5)$-invariant composite fields of the confined theory
should reproduce the global anomalies, $U(1)^3$, $U(1)^2 U(1)_R$,
etc of the original theory.
Denoting the fields of the confined theory by $X_i$, and their charges
under $G$ by $(q_i, r_i)$, we obtain four equations for the $q_i$'s
and $r_i$'s.
There is no simple solution to these equations.
Allowing only charges below 50, we need at least 5 fields
to obtain a solution.
We conclude then that the global symmetry is (probably) broken.
Since there are no classical flat directions, supersymmetry
is (probably) broken.

The supersymmetry breaking scale is proportional to the only scale
in the problem, which is the strong coupling scale of $SU(5)$.

\subsection{Direct analysis: the $3-2$ model}
\label{32}
The $3-2$ model is probably the canonical example
of supersymmetry breaking~\cite{Affleck:1984xz}.
It is certainly one of the simplest models in the sense
that it has a small gauge group $SU(3)\times SU(2)$,
and relatively small field content.
But it is actually not the simplest model to analyze.
Still, this makes it an interesting example, and we will
use it to demonstrate several important points.
We will see how the superpotential is determined
by holomorphy and symmetries. The basic observation
we will use is that the parameters of the theory can be
thought of the VEVs of background fields. 
The notion of holomorphy can then be extended to these parameters.

Furthermore, this model will also demonstrate the three types of analysis
detailed in the beginning of this section.
We will first establish supersymmetry breaking by the indirect argument we saw
in section~\ref{ftype}: we will show that the model has no flat directions
and a broken global symmetry.
We will then derive the exact superpotential of the theory and
show that it gives at least one non-zero $F$-term.
Finally, we will choose parameters such that the minimum
is calculable.

\subsubsection{Classical theory}
\label{classical}
The field content of the model is $Q\sim(3,2)$, $\bar{Q}_A\sim(\bar3,1)$,
 $L\sim(1,2)$ with $A=1,2$.
We add the superpotential 
\beq\label{wtree}
\wtree = \lambda Q\cdot \bar{Q}_2\cdot L\ .
\eeq
As explained in section~\ref{sscalar}, 
we should first find the $D$-flat directions,
and these can be parametrized by the classical gauge-invariants
that we can make out of the chiral fields
\beqa
X_A&=& Q\cdot \bar{Q}_A\cdot L = Q_{i\alpha} \bar{Q}_A^i L_\beta 
\epsilon^{\alpha\beta} \\
Y=\det(Q\cdot \bar{Q})&=& \epsilon^{\alpha\beta} \epsilon^{A B}\, 
(Q_{i\alpha} \bar{Q}_A^i)\, Q_{j\beta} \bar{Q}_B^j \ ,
\eeqa
where $i$ ($\alpha$) is the $SU(3)$ ($SU(2)$) gauge index. 
To see this, it is easy to start by making $SU(3)$ invariants: 
$Q_\alpha\cdot\bar{Q}_A$.
These are $SU(2)$ doublets, and together with the remaining
doublet $L_\alpha$, they can be combined into the $SU(2)$ invariants
$X_A$ and $Y$.  

Next, we should find the subspace of the $D$-flat directions
for which all $F$-terms vanish. 
Consider for example the requirement that  the $L$ $F$-term
vanishes,
\beq
{\partial W\over \partial L_\alpha}= \lambda Q_\alpha\cdot \bar{Q}_2 =0\ .
\eeq
Contracting this equation with $L_\alpha$ we see that $X_2=0$.
Similarly, you can show that $X_1=Y=0$.
Thus, there are no flat directions classically--only the origin
is a supersymmetric point.

Remembering our indirect criterion of section~\ref{ftype}, let's consider
the global symmetry of the model.
The only anomaly-free symmetry that's preserved by the 
superpotential~(\ref{wtree}), is $U(1)\times U(1)_R$,
with charges $Q(1/3,1)$, $\bar{Q}_1(-4/3,-8)$, $\bar{Q}_2(2/3,4)$,
and $L(-1,-3)$.
If we can show that this global symmetry is broken,
we'll know that supersymmetry is broken.

\subsubsection{Exact superpotential}
\label{exact}
So let's turn to the quantum theory. We already know that
only non-perturbative effects can change the potential (and
in particular ``lift'' the classical zero potential at the origin).
We also mentioned that the tricky part is to find the proper variables,
for which the K\"ahler potential is well behaved.
Our first task is then to find such variables and derive the 
superpotential~\cite{Intriligator:1996pu}.
Let's first see if we missed any gauge invariants.
The way we constructed the gauge invariants above was to contract
$SU(3)$ indices first. What happens if we do it the other way around?
We find one new gauge invariant
\beq
Z= (Q^2)\cdot (Q\cdot L) = \epsilon^{ijk} Q_{i\alpha} Q_{j\beta} 
\epsilon^{\alpha\beta} Q_{k\gamma} L_\delta \epsilon^{\gamma\delta}\ .
\eeq
Note that $Z=0$ classically.

We turn  now to deriving the superpotential. 
Beyond tree level, there can be contributions
to the superpotential generated by the $SU(3)$ and
$SU(2)$ dynamics.
To analyze these, it is useful to consider various limits.

Take first $\Lambda_3>>\Lambda_2$, and $\lambda$ much smaller
than the gauge couplings.
Then we have an $SU(3)$ theory with two flavors.
An $SU(3)$ instanton then gives rise to the superpotential
\beq\label{w3}
W_3={\Lambda_3^7\over Y}\ .
\eeq
Below we will see that~(\ref{w3}) is the most general
superpotential allowed by the symmetries of the theory.

But before doing that, let's note that we can already
conclude that supersymmetry is broken!
As a result of the the superpotential~(\ref{w3}), 
the ground state is at non-zero $Y$.
But $Y$ appears in the superpotential, so its $R$-charge must
be non-zero (you can check that it is indeed $-2$).
Therefore the global $R$-symmetry is broken, and since there
are no flat directions, supersymmetry must be broken too.

Note the difference between an $R$- and non-$R$ symmetry in
this respect.
We were able to conclude that the $R$ symmetry is broken
because a certain superpotential term is non-zero at
the ground state, and any superpotential term is
charged under the $R$-symmetry (assuming of course that there is
an $R$ symmetry that the superpotential preserves).
Since the superpotential is neutral under non-$R$ symmetries,
we cannot conclude analogously that a non-$R$ symmetry is broken.

Let us now show that the $SU(3)$ superpotential must
be of the form~(\ref{w3}).
In fact, we will show this more generally 
for an $SU(N)$ gauge theory with $F<N$ flavors $Q$ and $\bar{Q}$.
The global symmetry of this theory is 
$SU(F)_L\times SU(F)_R\times U(1)_B\times U(1)_R$,
with $Q\sim(F, 1, 1, (F-N)/F)$, and $\bar{Q}\sim(1,\bar{F},-1, (F-N)/F)$. 
The superpotential must be gauge invariant, so it can only
depend on the ``mesons'', $M_{ij}=Q_i\cdot \bar{Q}_j$
(with a slight abuse of notation, we are using now Latin indices
to denote both $SU(F)_L$ and $SU(F)_R$ indices, with $i,j=1,\ldots,F$).
So $W=W(M_{ij})$.

Furthermore, the superpotential better be invariant under 
$SU(F)_L\times SU(F)_R$, so $W=W(\det{M})$, where $M$ stands
for the meson matrix.
Now $\det{M}$ is neutral under $U(1)_B$, but has $U(1)_R$ charge
$2(F-N)$.
Therefore
\beq
W\propto \left( 1\over\det 
\bar{Q}\cdot Q\right)^{1\over N-F}\ .
\eeq
The only other thing $W$ can depend on is the $SU(N)$ scale
$\Lambda^{3N-f}$, so on dimensional grounds it is of the form
\beq\label{sun}
W= \mathrm{const} \left( \Lambda^{3N-F}\over\det 
\bar{Q}\cdot Q\right)^{1\over N-F}\ .
\eeq
Note that holomorphy was crucial in this argument---without it 
we could make invariants such as $Q^\dagger Q$.
Also note that we have just proven the non-renormalization theorem
for this theory. We did not put in any tree-level superpotential,
so $W_{tree}=0$. We argued that~(\ref{sun}) is the most general
form of the superpotential in the quantum theory.
But radiative corrections can only produce positive powers
of the fields. So indeed the tree-level superpotential is not corrected
radiatively.

Of course, we have only shown that the superpotential~(\ref{sun})
is allowed.
We haven't shown that it is actually generated, because
that's much harder~\cite{Affleck:1983mk,Cordes:1985um}.
But it is generated, by an instanton for $F=N-1$, and
by gaugino condensation for other $F<N$.
Going back to the $3-2$ model, an $SU(3)$ instanton generates
the superpotential~(\ref{w3}).

Finally, we get to the $SU(2)$ dynamics. 
In the limit $\Lambda_2\gg\Lambda_3$,
we have $SU(2)$ with two flavors.
The classical moduli space of this theory is parametrized
by the ``mesons'' $V_{ij}=Q_i\cdot Q_j$, $V_{i4}= Q_i\cdot L$.
An $SU(2)$ instanton modifies this moduli space, so that, at the
quantum level, the moduli space is given by the $V$'s subject
to the constraint 
\beq\label{w2}
W= A\, (\epsilon^{i_1 i_2 i_3 i_4} V_{i_1 i_2} V_{i3 i_4} 
-\Lambda_2^4) = A (Z-\Lambda_2^4)\ .
\eeq
where $A$ is a Lagrange multiplier.

We can now use these different limits to obtain the full
superpotential of the model, which is a function
\beq
W= W(X_A, Y, Z, \lambda, \Lambda_3^7, \Lambda_2^4)\ .
\eeq
As in the $SU(N)$ example above, we want to use
the global symmetry, which in this case is $U(1)\times U(1)_R$
to constrain this function.
However $\lambda$, $\Lambda_3$ and $\Lambda_2$ are of course neutral under this
symmetry, so that wouldn't work.
Note that in our $SU(N)$ example this was not a problem, because
there was only one parameter in the theory, $\Lambda$, and at the last
step we could constrain the way $\Lambda$ enters on dimensional grounds.
So we need symmetries under which $\lambda$, $\Lambda_i$ are charged,
i.e., global symmetries that are broken by the tree-level
superpotential, and/or have global anomalies.
In particular, we want to treat $\lambda$ as a background
field, or spurion, and use the fact that the superpotential
cannot depend on $\lambda^\dagger$.

The simplest symmetries to consider  the following:
Introduce $U(1)_Q$ under which $Q$ has charge 1, with all
other fields neutral. Under this symmetry, $\lambda$ has charge $-1$,
$\Lambda_3^7$ has charge 2, and $\Lambda_2^4$ has charge 3.
It is probably clear why $\lambda$ has charge $-1$.
We are introducing a ``fake'' symmetry and treating $\lambda$ as
a background field charged under this symmetry.
For the superpotential to be invariant under $U(1)_Q$,
$\lambda$ must have charge $-1$.
Let's now see why we can think of $\Lambda_3^7$ as having charge 2.
The $U(1)_Q$ symmetry is anomalous. Therefore, if we rotate
$Q$ by this symmetry, we will shift the $SU(3)$ $\theta$-angle.
The shift is proportional to the number of $SU(3)$ fermion zero modes
charged under the global symmetry. This number is 2, because $Q$ also
has an $SU(2)$ index.
Finally, recall that 
\beq
\Lambda^b=\mu^b e^{-{8\pi^2\over g^2(\mu)} + i\theta} \ ,
\eeq
so under the anomalous rotation, $\Lambda_3^7$ has charge 2.

\noindent
{\bf Exercise:} Introduce similarly $U(1)_{\bar{Q}_1}$,
$U(1)_{\bar{Q}_2}$, and $U(1)_L$, and compute the charges
of $\lambda$, $\Lambda_3^7$, $\Lambda_2^4$ under these symmetries.
Then use these symmetries, together with $U(1)_R$, to show
that the superpotential is of the form
\beq
W= {\Lambda_3^7\over Y}\, f(t_1, t_2) + A(Z-\Lambda_2^4)\, g(t_1,t_2)\ ,
\eeq
where $f$ and $g$ are general functions of
\beq
t_1={\lambda X_2 Y\over \Lambda_3^7}\ , \ \ \ \ t_2={Z\over\Lambda_2^4}\ .
\eeq

Now consider the limit 
\beq
X_2\,,\lambda_3\,,\lambda\rightarrow0\ .
\eeq
In this limit, $t_1$ and $t_2$ can take any value,
and we know
\beq
W\rightarrow A(Z-\Lambda_2^4)\ .
\eeq
Therefore $g(t_1,t_2)\equiv1$.
Now take 
\beq
Y\rightarrow\infty\,, \ \  \ \lambda\rightarrow0\ .
\eeq
Again $t_1$ and $t_2$ can take any value.
But for large $Y$ VEVs, the gauge symmetry is completely
Higgsed with the gauge bosons very heavy.
The low-energy theory is therefore weakly coupled,
and the superpotential is given by
\beq\label{w32}
W= {\Lambda_3^7\over Y} +\lambda X_2
\eeq
so that $f(t_1,t_2)=1+t_1$.
We then have the full superpotential
\beq
W= {\Lambda_3^7\over Y} + A(Z-\Lambda_2^4)+ \lambda X_2 \ .
\eeq
Since
\beq
{\partial W\over \partial X_2}\neq0\ ,
\eeq
supersymmetry is broken.

We assumed here that the K\"ahler potential is
non-singular in $X_2$.
This is justified because the theory is driven away from the origin
by the first term of~(\ref{w32}), so that the gauge symmetry is 
completely broken. We can then integrate out the heavy gauge bosons,
and the low energy theory can be described in terms of the gauge
invariants $X_A$, $Y$ and $Z$.
Note that, as a result, the tree-level superpotential becomes linear
in the fields, just as in the O'Raifeartaigh model.

Finally, we note that we derived the non-renormalization theorem
once again. The tree-level superpotential is not modified
by perturbative corrections.

\subsubsection{Calculable minimum}
\label{calc}
We established supersymmetry breaking by deriving the full
superpotential of the theory.
We can now choose parameters for which the minimum
is calculable.
For $\Lambda_3\gg \Lambda_2$, $\lambda\ll1$, $Y$ gets
a large VEV, and the gauge symmetry is completely broken.
Because of the superpotential~(\ref{w32}), $Z$ gets
mass and we can integrate it out\footnote{Because $Z$ vanishes
classically, the term $Z^\dagger Z$ in the K\"ahler potential
is suppressed by some power of $\Lambda_2/v$, where $v$ is the typical
VEV. Therefore the $Z$ mass is enhanced by $v/\Lambda_2$, and we
can indeed integrate it out.},
to get
\beq
W={\Lambda_3^7\over Y} +\lambda X_2\ .
\eeq
Since the theory is weakly coupled in this limit,
the K\"ahler potential is just the canonical K\"ahler
potential
\beq\label{kahler32}
Q^\dagger Q + \bar{Q}_A^\dagger \bar{Q}_A + L^\dagger L\ ,
\eeq
and we can calculate the potential, either in terms of
the elementary fields or in terms of the classical gauge invariants
$X_A$ and $Y$ (to use the latter, one needs to project~(\ref{kahler32})
on the classical moduli space).
In particular, it is easy to show that 
in terms of elementary fields, the typical VEV is
$v\sim \lambda^{-1/7}\Lambda_3$
and
$E_{vac}\sim \lambda^{5/14} \Lambda_3$.
This demonstrates the general features of calculable
minima mentioned at the beginning of this section.
As we lower the superpotential coupling $\lambda$,
the ground state is driven to large VEVs, for which the
theory is weakly coupled.
Note also that, as expected, the supersymmetry breaking scale
is proportional to to the relevant strong coupling scale,
($\Lambda_3$ in this limit) and to some positive power of
the Yukawa coupling $\lambda$.

We end this section with a few comments.

First, in this example, we were able to derive the exact
superpotential of the theory and conclude from it
that supersymmetry is broken.
It would have been much easier to just consider the
limit $\Lambda_3\gg\Lambda_2$, $\lambda\ll 1$,
and show that supersymmetry is broken as we did in 
section~\ref{calc}.
In general, even if we can only establish supersymmetry breaking
for some range of parameters, 
(say $\Lambda_3\gg\Lambda_2$, $\lambda\ll 1$),
we expect this to hold generally, because there
should not be any phase transition as we vary the parameters
of the theory.
However, the details of the breaking, such as the supersymmetry-breaking
scale, can be different.

Second, we used two examples to demonstrate the analysis
of supersymmetry breaking.
There is a long list of models that are known to break 
supersymmetry~\cite{Shadmi:1999jy}.
The analysis of these models involves many interesting ingredients
and phenomena: quantum removal of flat directions, 
supersymmetry breaking without $R$ symmetry, and the use of a Seiberg-dual
theory to establish supersymmetry breaking,
to name but a few.
Unfortunately, there is no fundamental organizing principle that would allow
us to systematically classify known models, or to guide us
in the quest for new ones.

\section{Mediating the breaking}
\label{mediating}
We now know that supersymmetry can be broken,
and that if broken dynamically, its scale
is proportional to some strong coupling scale, $\Lambda$,
which can be much lower than the Planck scale.
In fact, this is all we need from the previous
sections in order to discuss the mediation of supersymmetry
breaking to the MSSM.

The MSSM contains many soft supersymmetry-breaking 
terms:
scalar masses, gaugino masses, $A$-terms etc.
This is often cited as a drawback of supersymmetry.
But in any sensible theory, the soft terms must be
generated by some underlying theory, and this underlying theory
may have very few parameters. 
In fact, as we will see, if the soft terms are generated by anomaly-mediation,
they are controlled by a {\it single} new parameter---the overall supersymmetry
breaking scale.

The MSSM soft terms were discussed in detail in the lectures
of Wagner, Masiero and Nir~\cite{Nir:2005js}.
As we saw in these lectures,
the soft terms determine the
way we will observe supersymmetry in collider experiments,
and are severely constrained by flavor changing processes.
Here we will discuss several mechanisms for generating the
soft terms
\begin{itemize}
\item Mediation by Planck-suppressed higher-dimension operators
(a.k.a. ``gravity mediation'')
\item Anomaly mediated supersymmetry breaking (AMSB)
\item Gauge mediated supersymmetry breaking (GMSB)
\end{itemize}
We will focus on AMSB,
because it is always present, and  
because it is probably the most tricky.

Suppose then that the fundamental theory contains, in
addition to the MSSM, some fields and interactions that
break supersymmetry (these are usually referred to as
a supersymmetry breaking ``sector'', and the MSSM is sometimes
referred to as the ``visible sector'').
We can think of the supersymmetry breaking sector as the $3-2$ model,
or the $SU(5)$ model we saw above, or even as a model with tree-level
breaking, if we don't mind having very small parameters in the
Lagrangian.
The question is then: What do we need to do in order to communicate
supersymmetry breaking to the MSSM, namely, generate the MSSM soft terms?

\subsection{Mediating supersymmetry-breaking by Planck-suppressed operators}
\label{planck}
The short answer to this question is---nothing.
The effective field theory below the Planck scale
generically contains higher dimension operators
that are generated when heavy states with masses
of order the Planck scale are integrated out.
These higher dimension terms couple the MSSM fields
to the fields of the supersymmetry breaking sector.
Denoting the MSSM matter superfields by $Q_i$,
where $i$ is a generation index,  
and a field of the supersymmetry breaking sector
by $X$, the K\"ahler potential is then of the form
\beq\label{vh}
Q_i^\dagger Q_i + X^\dagger X + 
c_{ij} {1\over\mpl^2} X^\dagger X Q_i^\dagger Q_j + \cdots
\ ,
\eeq
where $c_{ij}$ are order-one coefficients.
If $X$ has a non-zero $F$-term, the last term of~(\ref{vh})
gives rise to scalar masses for the $Q$'s:
\beq
\left( m_{\tilde{Q}}^2\right)_{ij} = c_{ij}\, 
\left\vert{F_x\over \mpl}\right\vert^2\ . 
\eeq
For the scalar masses to be around the electroweak scale
we need
\beq
{F_x\over\mpl}\sim 100{\mathrm{GeV}}\ ,
\eeq
or $\sqrt{F_x}\sim 10^{11}$GeV. 
So it is very easy to generate the required scalar masses.
However, there is no reason for the coefficients $c_{ij}$ 
to be flavor blind.
The fundamental theory above the Planck scale is certainly not
flavor blind, because it must generate the fermion masses we observe.
Generically then, this mechanism, which is usually referred to as
``gravity mediation'', leads to large flavor changing neutral
currents. There are some solutions to this problem.
One solution, which we heard about in Nir's lecture, 
uses flavor symmetries, with different generation fields
transforming differently under the symmetry, leading to
``alignment'' of the fermion and sfermion mass matrices~\cite{nirseiberg}.

In fact, the name ``gravity-mediation'' is misleading,
because the mass terms are not generated by purely gravitational
interactions.
Instead, they are mediated by heavy string states
which couple to the MSSM and to supersymmetry-breaking fields
with unknown couplings.

Can we suppress these dangerous contributions to the
masses? One way to do this, is to suppress the coefficients
$c_{ij}$.
It is easy to do this if there are extra dimensions~\cite{rs}.
For example, if the MSSM is confined to a 3-brane,
and the supersymmetry breaking sector lives on a 
different 3-brane, separated by an extra dimension,
then tree-level couplings of the the two sectors are
exponentially suppressed, $c_{ij}\sim \exp(-M R)$,
where $R$ is the distance between the branes, and $M$ is
the mass of the heavy state that mediates the 
coupling. Such models are called sequestered models.
\footnote{The $c_{ij}$'s can be suppressed in 4d theories
too, using ``conformal sequestering''~\cite{lutysundrum}.}

Assume then that tree-level couplings of the MSSM and
supersymmetry breaking sector are negligible.
As it turns out, gravity automatically generates
soft masses for the MSSM fields through the scale anomaly
of the standard model. This time, the mediation of supersymmetry
breaking is purely gravitational.

\subsection{Anomaly mediated supersymmetry breaking}
\label{amsb}
As we said above, we are assuming that apart from the MSSM,
the theory contains a supersymmetry-breaking sector.
Therefore, as mentioned in section~\ref{local}, the scalar auxiliary field
of the supergravity multiplet  develops a non-zero VEV $F_\phi$.
The couplings of this auxiliary field to the MSSM are contained in
eqn~(\ref{phi}) which we repeat here for convenience
\beq\label{phia}
{\cal L} = \int d^2 \theta \Phi^3 W(Q) 
+ \int d^4\theta \Phi^\dagger \Phi K(Q^\dagger, e^V Q)
+ \int d^2\theta\, \tau \, W^\alpha W_\alpha\ .
\eeq
Here $Q$ denotes collectively the MSSM matter fields,
and $V$ stands for the MSSM gauge fields.
Note that because
\beq
\Phi=1 +F_\Phi\theta^2\ ,
\eeq
$F_\phi$ has dimension one. We could instead write it
as $F_\Phi=F/\mpl$, where $F$ is dimension-2 as usual.

At first sight, it seems that the non-zero $F_\Phi$ has no
effect on the MSSM fields, because we can rotate it away
by rescaling
\beq
Q\rightarrow \Phi^{-1} Q\ .
\eeq
Note however that this assumes that the superpotential
is trilinear in the fields, as is true for the MSSM
apart from the $\mu$ term.
If the superpotential contains a quadratic term then
the rescaling gives, schematically,
\beq
\int d^2\theta  \Phi^3 [Q^3 + M Q^2] \rightarrow \int d^2\theta 
[Q^3 + M \Phi Q^2]\ .
\eeq 
Thus, an explicit mass parameter would pick up one power of $\Phi$
\beq
M\rightarrow M\Phi =M (1+ F_\Phi \theta^2)\ .
\eeq
We will come back to this point often in the following.
But as we said above, the MSSM classical Lagrangian
is scale invariant---no mass parameter appears, and therefore
the non-zero $F_\Phi$ has no effect.

This scale invariance is lost of course when we include 
quantum effects.
The gauge and Yukawa couplings become scale dependent,
and the dependence is controlled by the relevant $\beta$ 
functions.
We now have an explicit mass scale---the cut-off scale
$\Lambda_{UV}$. As we saw above, this mass scale will
pick up powers of $\Phi$. 
Since the latter has a non-zero $\theta^2$ component,
we will obtain supersymmetry breaking masses
for the MSSM fields~\cite{rs,Giudice:1998xp}.

Consider first gaugino masses. These will come from
\beq\label{walpha}
\int d^2\theta {1\over4g^2({\mu\over\luv})
} W^\alpha W_\alpha 
\eeq
since $\luv$ is rescaled by $\Phi$ (the simplest
way to see this is to think of $\luv$ as the mass of
regulator fields),  
(\ref{walpha}) becomes
\beq
\int d^2\theta {1\over4g^2({\mu\over\luv\Phi})
} W^\alpha W_\alpha =
\int d^2\theta \left[{1\over4\guv^2} +
{b\over32\pi^2}\, \ln{\mu\over\luv\Phi}
\right] W^\alpha W_\alpha \ ,
\eeq
where $b$ is the one-loop $\beta$ function coefficient
for the gauge coupling.
Substituting~(\ref{phia}) and expanding in $\theta$,
we get
\beq
{1\over 4g^2(\mu)} W^\alpha W_\alpha \Big\vert_{\theta^2}
-{b\over32\pi^2} F_\Phi \lambda^\alpha\lambda_\alpha\ .
\eeq
The last term is a mass term for the gaugino.
Going to canonical normalization for the gaugino,
\beq\label{gaugino}
m_\lambda(\mu) = {b\over2\pi}\, \alpha(\mu) F_\Phi\ .
\eeq

\noindent
{\bf Exercise: scalar masses}. Repeat this analysis for the scalars.
Start from the K\"ahler potential
\beq
\int d^4\theta Z\left({\mu\over\luv}\right)\, Q^\dagger Q\ ,
\eeq
where $Z$ is the wave-function renormalization.
After the rescaling this becomes
\beq
\int d^4\theta Z\left({\mu\over\luv
(\Phi^\dagger\Phi)^{1/2}}\right)\, Q^\dagger Q\ ,
\eeq
(The combination $(\Phi^\dagger\Phi)^{1/2}$ appears
because $Z$ is real).
Expand this to obtain
\beqa\label{scalar}
m_0^2(\mu)&=& -{1\over4} {\partial\gamma(\mu)\over\partial\ln\mu}
\, \vert F_\Phi\vert^2 \nonumber\\
&=&
{1\over4} \left[{b_g\over2\pi} \alpha_g^2{\partial\gamma\over\partial\alpha_g}
+{b_\lambda\over2\pi} \alpha_\lambda^2 
{\partial\gamma\over\partial \alpha_\lambda}
\right]\, \vert F_\phi\vert^2
\eeqa  
where 
\beq
\gamma(\mu) = {\partial \ln Z(\mu)\over\partial\ln(\mu) }
\eeq
is the anomalous dimension, $\alpha_g=g^2/(4\pi)$,
$\alpha_\lambda=\lambda^2/(4\pi)$, $\lambda$ is a Yukawa
coupling, and $\beta_\lambda$ is its one-loop $\beta$ function
coefficient.

The AMSB masses (and $A$-terms, which can be derived similarly)
are determined by the MSSM couplings, beta functions, and anomalous
dimensions. 
The only new parameter that appears is $F_\Phi$, which sets
the overall scale.
Since gaugino masses are generated at one-loop, and scalar masses
squared are generated at two loops, the masses are comparable,
of order a loop factor times $F_\Phi$.
Furthermore, the scalar masses are largely generation blind.
Apart from third generation fields, for which flavor-changing
constraints are rather weak, the masses are dominated by gauge contributions.
Thus, FCNC's are not a problem.
Finally, the expressions~(\ref{gaugino}) and (\ref{scalar})
are valid at any scale, and in particular, at low energies.
Thus, AMSB is extremely elegant.
Unfortunately, it predicts negative masses-squared for the sleptons,
because $\beta_g<0$ for $SU(2)$ and $SU(3)$.

So minimal AMSB does not work. 
Furthermore,
we can already guess, from the fact that the soft terms
can be calculated directly at low energies, that it will
not be easy to modify them by introducing new physics
at some high scale.
We will explain this in detail in section~\ref{fix}.
But before doing that, let's pause to consider
gauge mediation.
We will then use the results of this section
together with the results of the next section
to tackle the question of ``fixing'' anomaly mediation
in section~\ref{fix}.

\subsection{Gauge mediated supersymmetry breaking} 
\label{gmsb}
In the last two sections, we assumed that the supersymmetry
breaking sector and the MSSM only couple indirectly,
either through higher-dimension operators, or
through the supersymmetry breaking VEV of the
supergravity multiplet.
In this section, we will instead extend the MSSM,
and couple it, mainly through gauge (but typically also through
Yukawa) interactions, to the supersymmetry breaking sector.
The main ingredient of gauge mediation are
new fields, that are charged under the standard-model gauge group,
and couple directly to the supersymmetry breaking sector,
so that they get supersymmetry-breaking mass splittings
at tree level.
These fields are usually called the ``messengers'' of supersymmetry 
breaking.
The MSSM scalars and gauginos obtain supersymmetry-breaking
mass splittings at the loop level, from diagrams with messengers
running in the loop.

We cannot go into detailed model building here.
Instead, we will concentrate on the simplest
set of messenger fields.
Furthermore, to simplify the discussion, we will
focus on the $SU(3)$ gauge interactions, and ignore $SU(2)\times U(1)$.
Our discussion can be  trivially extended to include these.

We then consider a ``vector-like'' pair of messengers, 
chiral superfields $Q_3$ and $\bar{Q}_3$, transforming 
as a 3 and $\bar3$ of $SU(3)$ 
respectively~\cite{Dine:1994vc,Dine:1995ag}.
The messengers couple to the supersymmetry-breaking
sector through the superpotential
\beq
W_{mess}=X Q_3 \bar{Q}_3\ ,
\eeq
where $X$ is a standard-model singlet, with a non-zero VEV,
$\langle X\rangle =M$ and $F$-term VEV, $\langle F_X\rangle =F\neq0$.
The $Q_3$, $\bar{Q}_3$ fermions then get mass $M$.
The scalar mass terms are of the form
\beq
M^2 \vert \tilde{Q}_3\vert^2 + M^2   \vert \tilde{\bar{Q}}_3\vert^2
+ \left( F \tilde{Q}_3 \tilde{\bar{Q}}_3 + h.c.\right)\ ,
\eeq
so that
\beq
m^2_{\tilde{Q}_3} = M^2\pm F\ .
\eeq
The gluinos then get mass at one loop (with the $Q_3$ scalar and fermion
running in the loop)
\beq\label{mino}
m_\lambda={\alpha_3\over4\pi}\, {F\over M} 
+ {\cal O}\left({F\over M^2}\right)^2\ ,
\eeq
and the squarks get masses at two loops,
\beq\label{msca}
m_0^2\sim \left({\alpha_3\over4\pi}\right)^2\,\left({F\over M}\right)^2
+ {\cal O}\left({F\over M^2}\right)^4\ ,
\eeq
We will see how to calculate these masses in the following.

The masses only depend on the SM gauge couplings, and are therefore
flavor blind, so that there are no FCNC's.
The gaugino and scalar masses are again comparable,
and given by a loop factor times $F/M$.
We therefore want $F/M$ to be around $10^4-10^5$GeV.
For $M$ lower than, roughly, 10$^{16}$GeV, the $\mpl$ suppressed
contributions we saw in section~\ref{planck} are negligible.
They would contribute soft masses of the order of $F/\mpl$,
at least two orders of magnitude below the gauge-mediated masses.
(The AMBS masses are smaller by a loop factor.) 

We will now see a nice trick~\cite{gr}
for calculating the GMSB soft masses, to leading order in
the supersymmetry breaking, $F/M^2$.
In the model we considered above, the masses are generated
when the messengers are integrated out at $\langle X\rangle =M$.
The effective theory for the gluinos below the messenger scale depends
on $M$ through the gauge coupling,
\beq
{\cal L} = \int d^2\theta {1\over 4g^2(\mu)} W^\alpha W_\alpha\ ,
\eeq
with
\beq
{1\over g^2(\mu) }= {b_H\over 8\pi^2}\ln{X\over \luv} 
+{b_L\over 8\pi^2}\ln{\mu\over X}\ ,
\eeq
where $b_H$ is the one-loop beta-function coefficient
above $M$ (MSSM $+ Q_3 +\bar{Q}_3$) and $b_L$ is the one-loop
beta function coefficient below $M$ (MSSM).
The key point is that we promoted the VEV of $X$ to the
field $X$. 
Since $X=M + \theta^2 F$ the situation is completely analogous
to what we had in the previous section.
We can Taylor expand in $\theta$ to get the gaugino mass
\beq
m_\lambda(\mu)= {\alpha(\mu)\over4\pi} (b_L- b_H) {F\over M}\ .
\eeq
In our case, $b_L-b_H=1$, so we recover~(\ref{mino}).
Similarly, starting with the quark kinetic term we can
essentially repeat the derivation of scalar masses in AMSB,
to get~(\ref{msca}).

This concludes our short review of gauge mediation.

\subsection{How NOT to fix AMSB}
\label{fix}
As we saw above, minimal AMSB gives rise to tachyonic sleptons.
One might try to modify the slepton masses by adding some new 
physics at a high scale.
We will now show that this has no effect on the masses at
low scales.
We assume here that the only source of supersymmetry breaking 
in the visible sector is anomaly mediation.
For simplicity let us take the new fields to be the vector-like
pair $Q_3$, $\bar{Q}_3$ of the previous section.
We also add the superpotential 
\beq\label{mq3}
W=M Q_3 \bar{Q}_3\ .
\eeq
Now let us calculate the AMSB masses at low energies
below $M$.
For simplicity, we will consider gaugino masses only,
but a similar discussion applies for scalar masses
and $A$ terms.
Just above the scale $M$, the gaugino masses are given by
the usual AMSB prediction~(\ref{gaugino})
\beq
m_\lambda(\mu) = {b_H\over2\pi}\, \alpha(\mu) F_\Phi\ \ \ 
\ \ \ \ \ \mathrm{for} \ \ \ \mu > M\ ,
\eeq
where $b_H$ is the beta function coefficient for the
MSSM + $Q_3$, $\bar{Q}_3$, 
$$b_H= b_{\mathrm{MSSM}}+1\ .$$
At the scale $M$, we need to integrate out the heavy fields.
But because the superpotential~(\ref{mq3}) contains
an explicit mass parameter, these fields get supersymmetry-breaking
mass splittings at tree-level
\beq\label{mq3a}
W=M Q_3 \bar{Q}_3 \rightarrow \Phi M Q_3 \bar{Q}_3
= (1+ F_\Phi \theta^2) M Q_3 \bar{Q}_3
\ ,
\eeq
with the fermions at $M$, and the scalars at
$m^2= M^2\pm M F_\phi$. 
So $Q_3$ and $\bar{Q}_3$ behave just like the messengers
of gauge mediation!
We can calculate their contribution to the gaugino masses
just as we did in the previous section.
Clearly, the effect of this contribution is to precisely
cancel the $Q_3$ $\bar{Q}_3$ part in $b_H$,
so that below $M$, the gaugino mass is
\beq
m_\lambda(\mu) = {b_{\mathrm{MSSM}}\over2\pi}\, \alpha(\mu) F_\Phi\ \ \ 
\mathrm{for} \ \ \mu<M\ ,
\eeq
as in minimal AMSB.
The heavy fields decouple completely and have no effect
on the soft masses~\cite{rs,pr,kss}.

Note that it was crucial here that the new fields
 get mass in a supersymmetric manner.
To emphasize this, let's give an even simpler argument
for the decoupling.
Consider the low-energy theory below $M$,
\beq
\int d^2\theta\, \tau(\mu, M, \luv)\, W^\alpha W_\alpha \ .
\eeq 
On dimensional grounds
\beq
\tau=\tau\left({\mu\over\luv},{M\over\luv}  \right) \ .
\eeq
Rescaling explicit mass scales by $\Phi$
\beq
\tau\left({\mu\over\luv},{M\over\luv}  \right) \rightarrow
\tau\left({\mu\over\luv\Phi},{M\Phi\over\luv\Phi}  \right)=
\tau\left({\mu\over\luv\Phi},{M\over\luv}  \right)\ .
\eeq
The $\Phi$ dependence cancels out completely in $M$ so we recover
the minimal AMSB prediction.
Note that the cancellation only holds to leading order in $F_\Phi$.
The reason is that the AMSB masses are given fully by~(\ref{gaugino})
and~(\ref{scalar}), with no corrections at higher order in $F_\Phi$.
In contrast, the ``GMSB'' contributions from integrating out
$Q_3$ and $\bar{Q}_3$, do contain higher order corrections,
that are not captured by the trick we saw in the previous
section.

The same discussion applies to different heavy thresholds,
and in particular those associated with $D$ terms, which 
have attracted some attention lately~\cite{Harnik:2002et}.
The basic idea~\cite{kss} is to get slepton masses by adding a new $U(1)$
symmetry, under which the MSSM matter fields are charged.
Probably the simplest model~\cite{kss} involves new fields 
$h_\pm$, $\xi_\pm$, with charges $\pm1$ under the $U(1)$,
as well as gauge singlets $n_i$, $i=1,2$, and $S$,
with the superpotential
\beq
W= S(\lambda h_+ h_- -M^2) + y_1 n_1 h_+ \xi_-
+  y_2 n_2 h_- \xi_+\ .
\eeq
Because of the first term, $h_+$ and $h_-$ obtain VEVs
and break the $U(1)$.
All new fields get mass either by the Higgs mechanism
or through the superpotential.
With no supersymmetry breaking,  $h_+$ and $h_-$ 
get equal VEVs.
However, assuming that there is some supersymmetry breaking
sector, all fields get supersymmetry breaking masses
through AMSB.
In particular, for $y_1\neq y_2$, $h_1$ and $h_2$
have different soft masses and therefore different VEVs,
so that the $U(1)$ $D$-term is non-zero.
If the sleptons are charged under the $U(1)$, one
might naively think that the $D$ term affects the
slepton masses. 
But as explained in~\cite{kss}, this is not the case.
The model described above has no effect on the soft masses at
low energy, to leading order in the supersymmetry breaking,
$F/M^2$.
In~\cite{kss}, the surviving $F^4/M^2$ contributions were used in
order to generate acceptable slepton masses, using the
fact that these enter scalar masses-squared at {\it one}-loop.
The scale $M$ was generated dynamically from $F$, so that
it was roughly two orders of magnitude (an inverse loop-factor)
above $F$.

To conclude, we see that we cannot modify the AMSB predictions
at leading order in $F_\Phi$
using new heavy thresholds that get mass in the limit of 
unbroken supersymmetry.
Clearly then, there are two possible approaches
to fixing AMSB models: One is to use higher order
terms in the supersymmetry breaking $F_\Phi$.
The second is to introduce thresholds for which
some fields remain light in the limit of unbroken supersymmetry.

\vskip 0.15in
 \noindent{ \bf Acknowledgements}
\vskip 0.125in
\noindent
I thank the organizers, Stephane Lavignac and Dmitri Kazakov,
for running such a smooth and enjoyable school.
And I thank the students, who asked many good questions,
and made giving these lectures fun.


\end{document}